\documentclass{raa}
\usepackage{graphicx,times} 
\usepackage{natbib}
\usepackage{amssymb,amsmath}
\bibpunct{(}{)}{;}{a}{}{,}

\usepackage{soul,xcolor}
\setstcolor{red}

\usepackage[a4paper=true,pagebackref=true]{hyperref}
\hypersetup{pdftitle = The title of my PDF, pdfauthor = My name, pdfsubject= The subject, pdfkeywords = keyword1 keyword2 keyword3}
\hypersetup{colorlinks = true, linkcolor = green, anchorcolor = red, citecolor = blue, filecolor = red, pagecolor = red, urlcolor = red}
\begin{document}

\title{New continuum and polarization observations of the Cygnus Loop with FAST} \subtitle{II. Images and analyses}
\volnopage{Vol.0 (20xx) No.0, 000--000}
\setcounter{page}{1}

\author{Xiao-Hui Sun\inst{1}, Xu-Yang Gao\inst{2, 3}, Wolfgang Reich\inst{4}, Peng Jiang\inst{2, 3}, Di Li\inst{2,3}, Huirong Yan\inst{5, 6}, Xiang-Hua Li\inst{1}}

\institute{School of Physics and Astronomy, Yunnan University, Kunming 650500, China; {\it xhsun@ynu.edu.cn, xhli@ynu.edu.cn}\\
\and
National Astronomical Observatories, Chinese Academy Sciences, 20A Datun Road, Chaoyang District, Beijing 100101, China; {\it xygao@nao.cas.cn}\\
\and
CAS Key Laboratory of FAST, National Astronomical Observatories, Chinese Academy of Sciences, Beijing 100101, China
\and
Max-Planck-Institut f\"ur Radioastronomie, Auf dem H\"ugel, 53121 Bonn, Germany\\
\and
Institut f\"ur Physik und Astronomie, Universit\"at Potsdam, Haus 28, Karl-Liebknecht-Str. 24/25, D-14476 Potsdam, Germany\\
\and
Deutsches Elektronen-Synchrotron DESY, Platanenallee 6, D-15738 Zeuthen, Germany\\
\vs\no
{\small Received~~20xx month day; accepted~~20xx~~month day}
}

\abstract{We present total-intensity and polarized-intensity images of the Cygnus Loop supernova remnant (SNR) observed by the Five-hundred-meter Aperture Spherical radio Telescope (FAST). The high angular-resolution and high-sensitivity images enable us to thoroughly compare the properties of the northern part with the southern part of the SNR. The central filament in the northern part and the southern part have a similar foreground rotation measure, meaning their distances are likely similar. The polarization analysis indicates that the random magnetic field is larger than the regular field in the northern part, but negligible in the southern part. The total-intensity image is decomposed into components of various angular scales, and the brightness-temperature spectral index of the shell structures in the northern part is similar to that in the southern part in the component images. All these evidence suggest that the northern and southern part of the Cygnus Loop are situated and thus evolved in different environments of interstellar medium, while belonging to the same SNR.  
\keywords{ISM: supernova remnants --- ISM: magnetic fields --- polarization --- techniques: polarimetric}
}

\authorrunning{Sun et al.}
\titlerunning{Cygnus Loop with FAST}
\maketitle

\section{Introduction}

The Cygnus Loop (G74.0$-$8.5) is a prominent supernova remnant (SNR), which has been observed across a broad electromagnetic spectrum from radio to $\gamma$-rays. However, there is no consensus yet on the interpretation and physical nature of the Cygnus Loop. Particularly, the question on whether the southern part is a blowout region from the larger northern part or a separate SNR remains unclear~\citep{uyaniker+02, fesen+18}. 

In optical band, the Cygnus Loop clearly exhibits a Balmer H$\alpha$ shell delineating the boundary of the SNR~\citep{Levenson+98}. This thin shell traces the non-radiative shock front at a velocity close to 400~km~s$^{-1}$~\citep{fesen+18}, implying that the SNR is at the early stage of the adiabatic phase. The soft X-ray emission appears to be confined by the Balmer shell, but only fills the interior of the northern part of the SNR~\citep{aschenbach+leahy+99}. Strong [S~{\scriptsize II}] and [O~{\scriptsize III}]~lines as well as UV emission were detected towards the northeastern and the western parts of the SNR, adjacent to the Balmer filament~\citep{Levenson+98, fesen+18}, which indicates the interaction between shocks and interstellar clouds. Consequently, the shocks are transitioned to be radiative at a velocity around 100~km~s$^{-1}$~\citep{fesen+18}.

In radio band, the Cygnus Loop is composed of two overlapping shell structures with centers aligned roughly in the north-south direction~\citep{uyaniker+04}. The northern radio shell corresponds well to the optical emission towards the north east and the west of the SNR. The strong radio emission from the southern shell, however, has no correspondence in other wavelengths. The two shells share a similar total-intensity spectrum~\citep{uyaniker+04, loru+21}. The polarization characteristics of the two shells are totally different, which can be readily interpreted by the scenario of two separate SNRs \citep{uyaniker+02, sun+06, west+16}.

The two-SNR scenario does not seem to be corroborated by observations at other wavelengths. The X-ray observations by \citet{uchida+08} indicate that the plasma temperature and the abundances of Ne, Mg and other elements are similar between the northern and southern parts of the Cygnus Loop, which supports that the southern shell is just a blowout region as proposed by \citet{aschenbach+leahy+99}. Optical observations by \citet{fesen+18} reveal that the morphology of the southern shell resembles that of a second possible blowout region in the Cygnus Loop. \citet{katsuda12} reported the discovery of an X-ray pulsar and a pulsar-wind nebula in the southern part of the SNR, which would prove the southern shell to be an individual SNR. However, \citet{Halpern+Gotthelf19} identified these sources as a Seyfert galaxy and a cluster of galaxies, respectively, and a pulsar associated with the Cygnus Loop thus remains elusive.

For the blowout scenario, the exact process is uncertain. The simulations by \citet{fang+17} demonstrated that a supernova explosion in a cavity evacuated by the winds from the progenitor star can reproduce the morphology of the Cygnus Loop. A multi-wavelength analysis by \citet{fesen+18} suggested that the Cygnus Loop lies in an extended low-density region rather than a wind-driven cavity, and the interaction between the shocks and the interstellar clouds causes the morphology. In contrast, \citet{tenorio+85} proposed that the supernova explosion occurred in a dense molecular cloud in the southern shell, and the northern part of the SNR is the result of a blowout. However, there is no indication of a dense molecular cloud surrounding the southern shell, as can be seen from the multi-wavelength images presented by \citet{fesen+18}. \citet{meyer+15} put forward an alternative explanation where a massive runaway star shaped the interstellar medium with a bow shock and a supernova explosion afterwards in the shaped region resulted in the morphology. 

We conducted new continuum and polarization observations of the Cygnus Loop with the Five-hundred-meter Aperture Spherical radio Telescope~\citep[FAST, ][]{nan+11, jiang+20} to investigate the properties of the SNR. In Paper I~\citep{sun+21}, we verified the data processing and demonstrated the exceptional imaging capability of FAST. In this paper, we will present a detailed analysis of the Cygnus Loop. The paper is organized as follows. We describe the observations and results in Sect.~\ref{sect:obs}, present discussions in Sect.~\ref{sect:discussions}, and summarize the conclusions in Sect.~\ref{sect:conclusion}. 

\section{Observations and results}
\label{sect:obs}

We presented the images from the scanning observations in right ascension (RA) direction (FAST project code: 2019a-125-C) in Paper I. There are stripes along the scanning directions in those images, which are caused by drifting of the system. In order to eliminate the stripes, we conducted new scanning observations in declination (Dec) direction (FAST project code: PT2021\_0111), and combined the observations in both directions to produce the final maps.

The data processing was described in Paper I. In summary: (1) radio frequency interference (RFI) and H~{\scriptsize I} line were mitigated; (2) leakage was corrected and the antenna-temperature scale was established according to the injected reference signal; (3) gains for all the beams were derived from the calibrators 3C~138 or 3C~48; (4) gains were applied to the scans; (5) maps were constructed by combining all the scans, and were multiplied by the conversion factors from antenna temperature to main-beam brightness temperature ($T_{\rm b}$). These procedures were repeated for each individual frequency and scan direction. The final maps were obtained by combining the maps from the two orthogonal directions with the basket-weaving method by \citet{emerson+88}.

After processing all the observations, we obtained frequency cubes of Stokes parameter $I$, $Q$, and $U$ containing nearly 30\,000 frequency channels. The width of each frequency channel is about 7.63~kHz. We smoothed all the images to a common angular resolution of $4\arcmin$.

\subsection{Total intensity}

\begin{figure}
    \centering
    \includegraphics[width=0.9\textwidth]{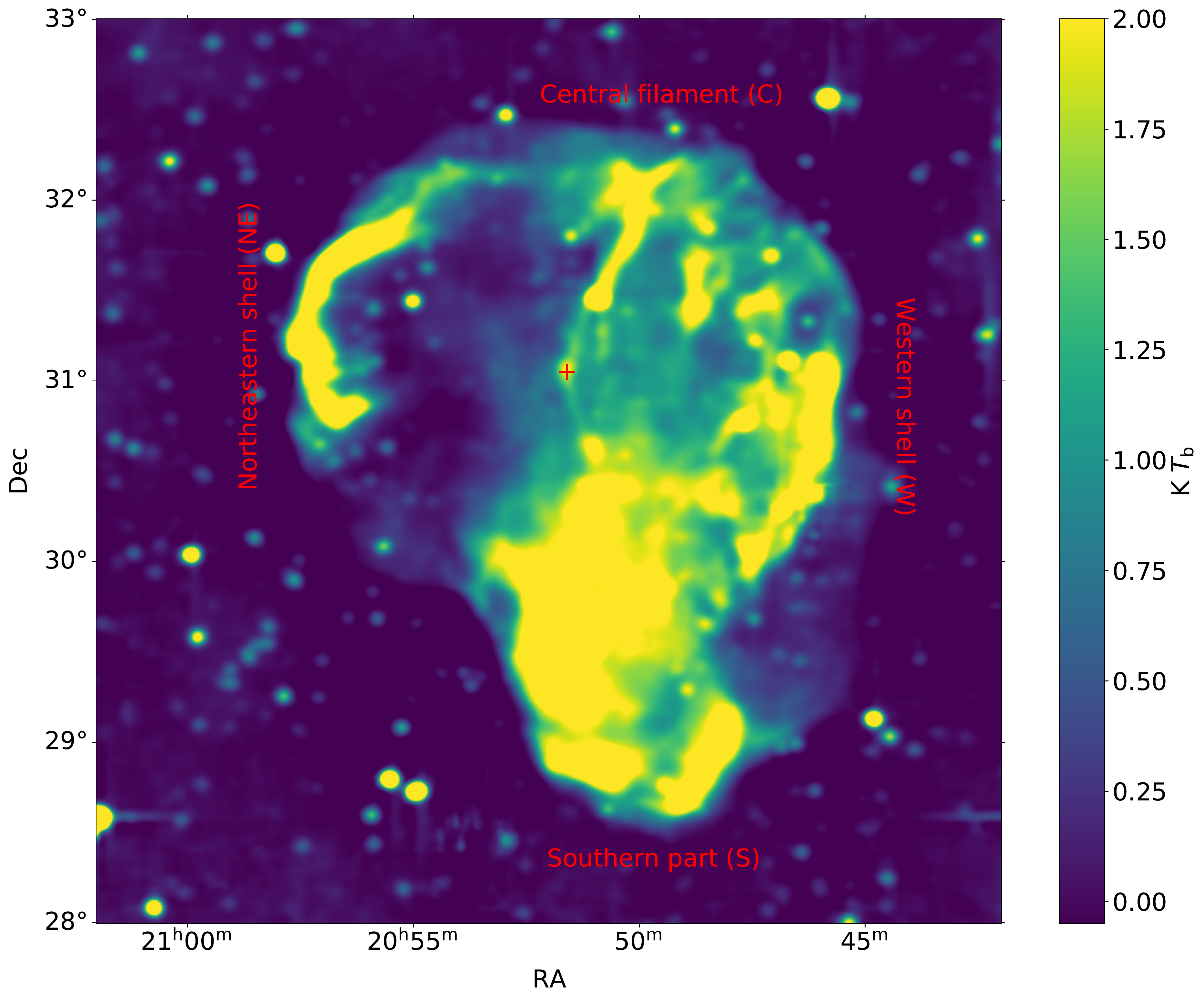}
    \caption{Total-intensity image of the Cygnus Loop from FAST at 1.28~GHz and 4$\arcmin$ angular resolution. The names of some features are marked. The geometric enter of the northern part is marked with a plus sign.}
    \label{fig:cyg_i}
\end{figure}

We averaged the frequency cube of $I$ by taking the median values and obtained the total-intensity image centered at about 1.28~GHz, as shown in Fig.~\ref{fig:cyg_i}. In comparison with the image in Paper I, most of the scanning effects have been successfully removed. There is strong emission from the northern part of the SNR consisting of the northeastern shell, the central filament, and the western shell, and the southern part of the SNR, which is similar to the previous radio observations~\citep{uyaniker+04, sun+06}. The northern part has a well-defined circular shape with the geometric center roughly at $(20^{\rm h}51^{\rm m}36^{\rm s},\,31\degr03\arcmin)$, marked as a plus sign in Fig.~\ref{fig:cyg_i}.

The rms noise measured from the map is about 16~mK~$T_{\rm b}$, lower than the value of 20~mK~$T_{\rm b}$ obtained from the map with only RA scans presented in Paper I. As discussed in Paper I, this rms noise is also much lower than the confusion level of about 34~mK~$T_{\rm b}$ estimated following \citet{condon+74} and \citet{meyers+17}, which seems to be an overestimate. However, the current value is 
consistent with the result published by \citet{uyaniker+99}.

\subsection{Polarization}

We applied the rotation measure (RM) synthesis~\citep{brentjens+05} and RM clean~\citep{heald+09} to the frequency cubes of $Q$ and $U$ to obtain Faraday depth ($\phi$) cubes of complex polarized intensity, defined as Faraday dispersion function $F(\phi)$. The details of the process refer to Paper I. We then searched for peaks in $|F(\phi)|$ towards each individual pixel of the maps. The peak value of $|F(\phi)|$ and the peak location $\phi$ correspond to the polarized intensity and RM, respectively. Note that RM in rad~m$^{-2}$ can be related to thermal electron density $n_e$ in cm$^{-3}$ and line-of-sight magnetic field $B_\parallel$ in $\mu$G as,
\begin{equation}\label{eq:rm}
   {\rm RM}=0.81\int_{\rm source}^{\rm observer} n_e B_\parallel {\rm d}l
\end{equation}
where ${\rm d}l$ is the increment of path length along line of sight in unit of pc. 

\begin{figure}
    \centering
    \includegraphics[width=0.8\textwidth]{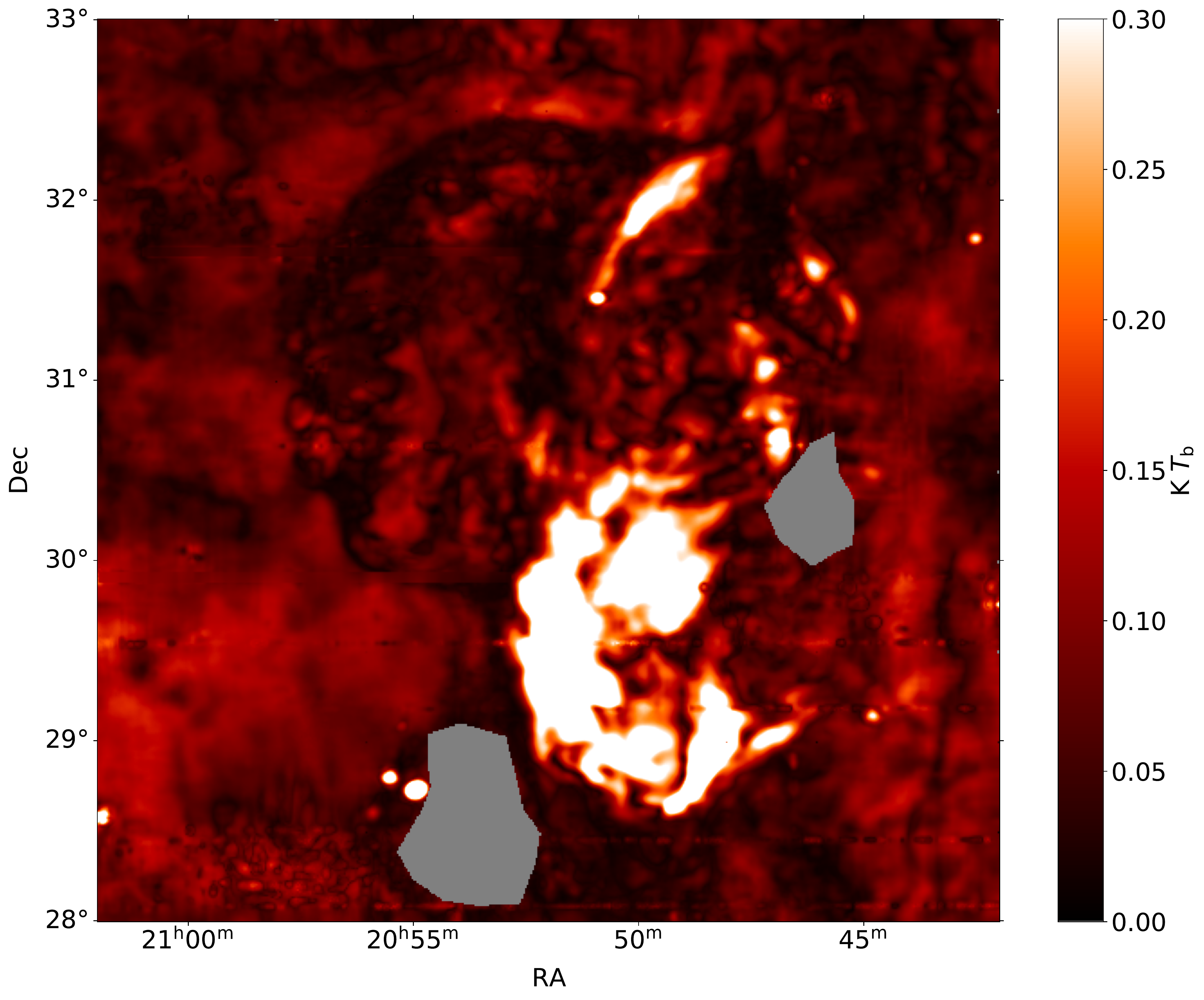}
    \includegraphics[width=0.8\textwidth]{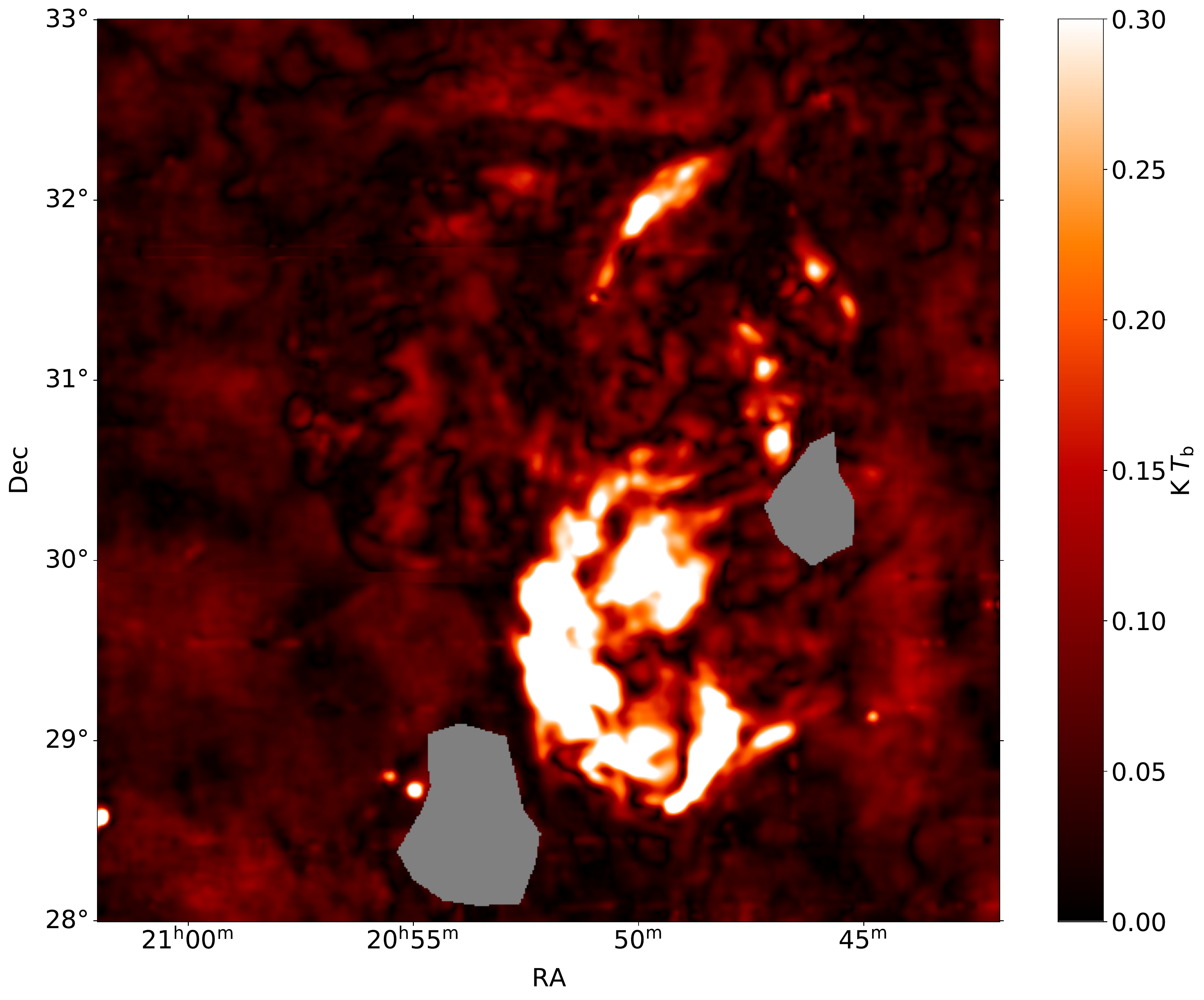}
    \caption{Polarized-intensity images of the Cygnus Loop from FAST at 1.28~GHz, derived from RM synthesis (top panel) and calculated directly from $Q$ and $U$ (bottom panel). The two areas with ``damaged" data are marked in grey.}
    \label{fig:cyg_pi}
\end{figure}

We show the polarized-intensity map in Fig.~\ref{fig:cyg_pi} (top panel). We also calculated polarized intensity directly from the averaged $Q$ and $U$ as $PI=\sqrt{Q^2+U^2}$, and the result is also shown in Fig.~\ref{fig:cyg_pi} (bottom panel). The two areas centered at about RA = 20$^{\rm h}$54$^{\rm m}$, Dec = 28$\fdg5$ and RA = 20$^{\rm h}$46$^{\rm m}$, Dec = 30$\fdg2$ with damaged data are marked as blank. As can be seen from these images, the polarized intensities derived with both methods are nearly identical. For some regions such as the source at the end of the central filament, the polarized intensity calculated directly from $Q$ and $U$ is much weaker, indicating depolarization after averaging $Q$ and $U$ over all the frequency channels. We therefore used the polarized intensity from RM synthesis for the analysis in below. 

The rms noise for polarized intensity is about 4~mK~$T_{\rm b}$, which is the same as that presented in Paper I. The rms noise does not decrease by adding the Dec scans because of the fluctuations of Galactic diffuse polarized emission. There are several RA stripes at declination of about $28\degr27\arcmin$, $29\degr32\arcmin$, and $31\degr42\arcmin$, which are caused by bad scans.

There is a stark difference for polarization between the northern and southern parts. Towards the north, only the central filament and several fragments from the western shell show polarized emission and the rest is completely depolarized. Towards the south, the polarized emission closely follows the total intensity for the whole area. This is consistent with previous observations presented by \citet{uyaniker+04} and \citet{sun+06}.    

\begin{figure}
    \centering
    \includegraphics[width=0.8\textwidth]{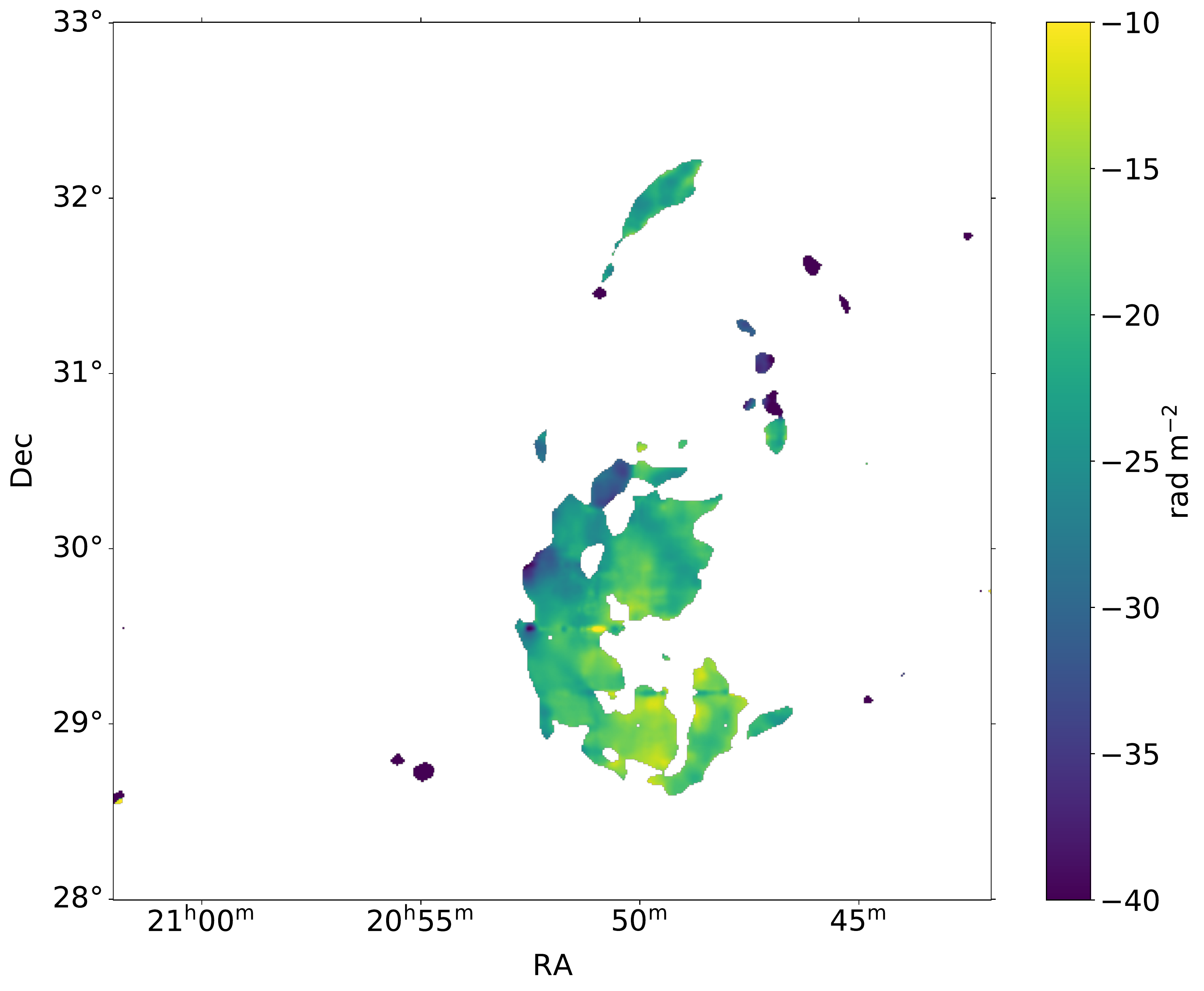}
    \caption{RM map of the Cygnus Loop.}
    \label{fig:rm}
\end{figure}

The RM map is shown in Fig~\ref{fig:rm}. RMs were measured towards the central filament, the southern part and some extragalactic sources, showing strong polarized emission. For the southern part of the SNR, the mean RM is about $-16$~rad~m$^{-2}$ for the lower part and about $-20$~rad~m$^{-2}$ for the upper part, consistent with the value of $-$21~rad~m$^{-2}$ derived by \citet{sun+06} based on data at 4800~MHz and 2695~MHz. For the central filament, the average RM is about $-$21~rad~m$^{-2}$, very similar to that of the southern part. Towards all these areas, the RM distribution is very smooth with a standard deviation of about 3~rad~m$^{-2}$.

\section{Discussions}
\label{sect:discussions}

\subsection{A multi-wavelength view}
\begin{figure}
    \centering
    \includegraphics[width=0.8\textwidth]{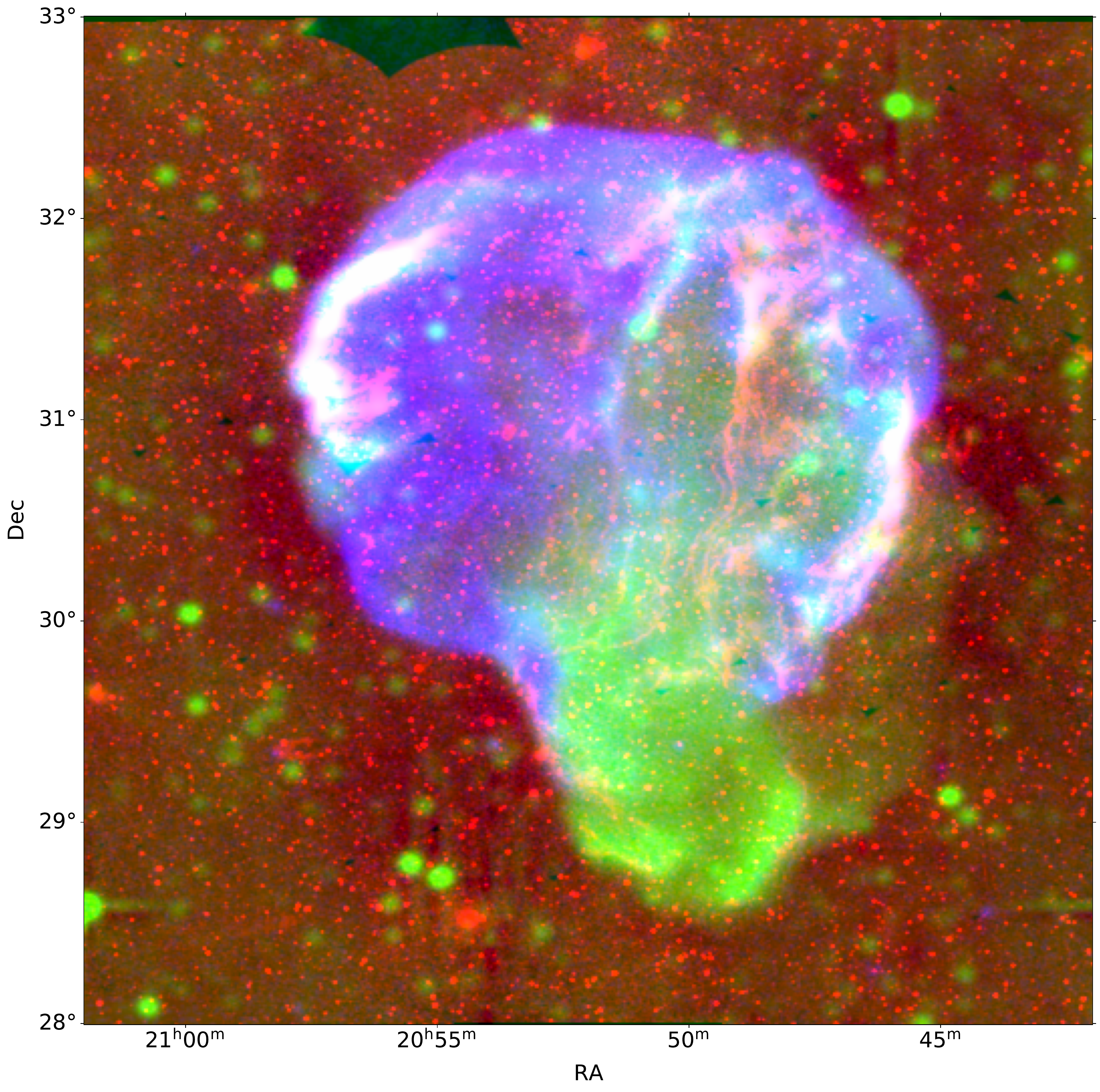}
    \includegraphics[width=0.8\textwidth]{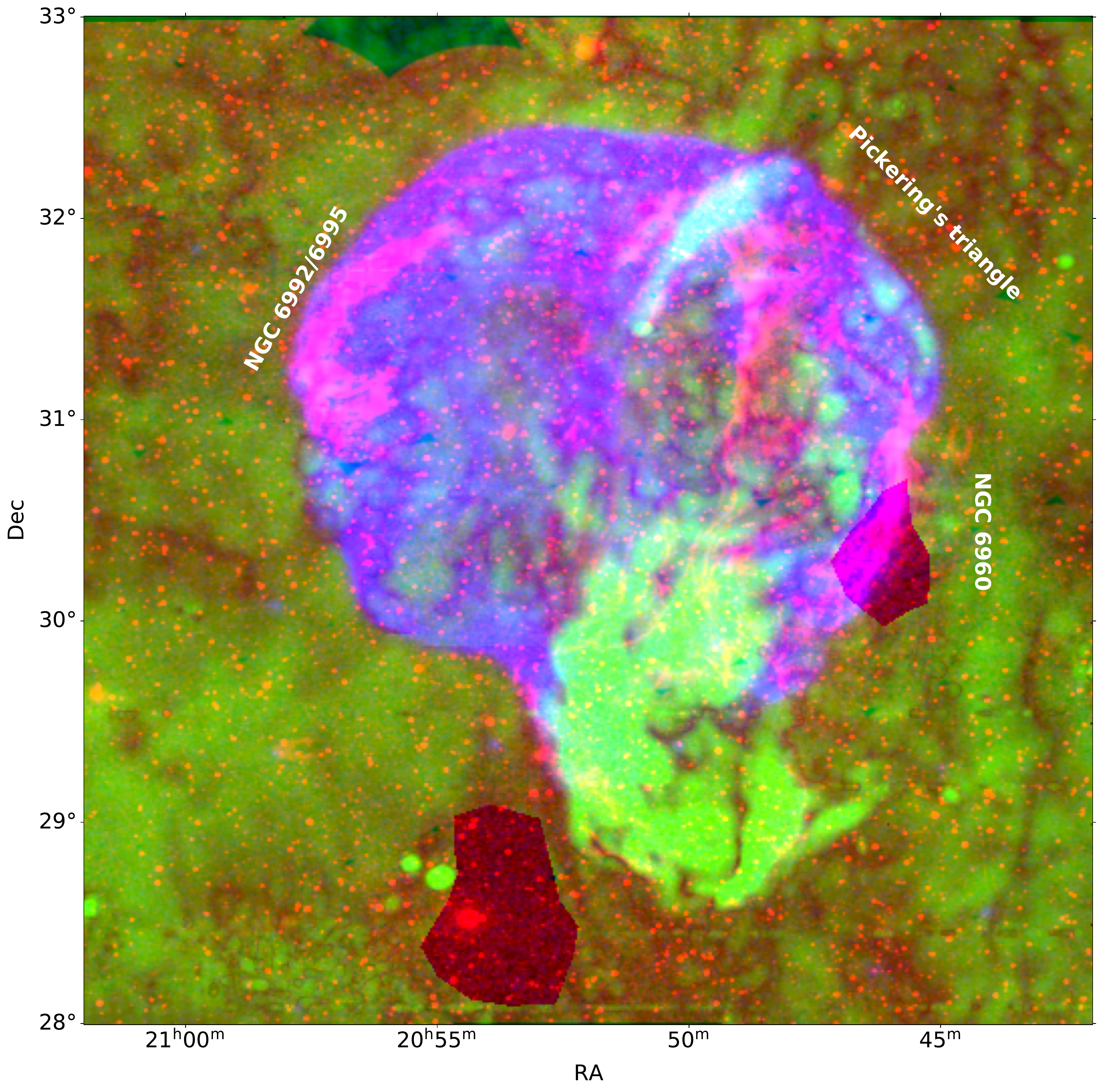}
    \caption{RGB images of the Cygnus Loop. Red: GALEX near UV, Green: FAST total intensity (top panel) and polarized intensity (bottom panel), Blue: ROSAT X-ray.}
    \label{fig:cyg_rgb}
\end{figure}

We retrieved ROSAT X-ray~\citep{rosat+99} and GALEX near UV~\citep{galexuv+17} images from SkyView~\footnote{https://skyview.gsfc.nasa.gov/current/cgi/titlepage.pl}, and combined them with the FAST radio data into RGB images shown in Fig.~\ref{fig:cyg_rgb}. Here the near UV data are in red, the radio total intensity and polarization data are in green, and the X-ray data are in blue. Emission visible in all three bands will appear in white. Similar RGB images have been presented by \citet{uyaniker+reich+04} and \citet{west+16} to compare the emission at different wavelengths.

The morphology of the Cygnus Loop in the three bands exhibits clear discrepancy. The X-ray emission is distributed across the entire northern part, with enhancements towards the edges outlining the boundary of the SNR. The near UV emission is only concentrated on the northeastern shell, the western shell, and the filament in the west of the radio central filament, corresponding to the optically identified bright nebulae (Fig.~\ref{fig:cyg_rgb}): NGC 6992/6995, NGC 6960, and the Pickering's triangle~\citep{fesen+18}. Both the X-ray and near UV emission are confined to the northern part with a weak extension towards the south. The radio total intensity is well correspondent with near UV and X-ray emission towards the northeastern shell and the western shell. Radio emission dominates the southern part. 

The northern part and the southern part of the Cygnus Loop have certainly undergone different paths of evolution, which resulted in the different appearance in all the three bands. Towards the north, the strong near UV emission tracing the interaction between blast wave and dense interstellar clumps~\citep{fesen+18}, is a signature of the SNR entering radiative phase~\citep[e.g.][]{dubner+15}. The center-filled X-ray emission is from the swept material or SN ejecta heated by the reverse shock. Towards the south, there is no clear indication of interactions with the ambient medium and the radio morphology resembles a typical SNR in adiabatic phase.   

\subsection{RM}
RM consists of foreground contribution from the magnetized thermal medium in front of the SNR and the intrinsic contribution from the medium inside the SNR. Towards the central filament and the southern part of the SNR, the degree of polarization, defined as the ratio between the polarized intensity and the total intensity, is similar at 1.4, 2.7, and 4.8~GHz, indicating little depolarization~\citep{sun+06}. This means that the RM contributed by the medium inside the SNR is negligible for these areas, otherwise depolarization towards lower frequencies (1.4~GHz) would be expected. 

We can estimate the foreground RM based on Eq.~\ref{eq:rm}, ${\rm RM}\sim0.81 n_e B_\parallel d$, where $d$ is the distance in pc. Using RMs and dispersion measures of pulsars by \citet{han+06}, the local magnetic field is about 2~$\mu$G pointing away from the Sun towards the Cygnus Loop. Based on the thermal electron-density model by \citet{yao+17}, the average electron density in the direction of the Cygnus Loop is about 0.015~cm$^{-3}$. Substituting these values yields an RM of about $-$18~rad~m$^{-2}$, consist with the RMs obtained for the central filament and the southern part. We also simulated the RM with the 3D emission models of the Galaxy by \citet{sun+10}, and derived an RM of about $-$13~rad~m$^{-2}$, consistent with the RM of the lower section of the southern part. These estimates confirm that the RMs are mainly from the foreground contribution. 

The RMs of the central filament and the southern part of the Cygnus Loop represent the contribution of the foreground medium. These RMs are consistent with each other within the errors given the scattering of about 3~rad~m$^{-2}$. Therefore the similar RM values suggest that these areas should likely be located at the similar distance. 

\subsection{Depolarization}
The bright northeastern shell and the western shell in total intensity (Fig.~\ref{fig:cyg_i}) do not have correspondence in polarized intensity (Fig.~\ref{fig:cyg_pi}). However, polarized emission from these two areas was detected at both 2.7~GHz and 4.8~GHz~\citep{sun+06}, implying a complete depolarization at 1.4~GHz. The low-polarization area extends from the northeastern shell and forms a circular region with a sharp boundary, as can be seen from Fig.~\ref{fig:cyg_pi}. Beyond the boundary, there are extended patches with low-level polarized emission that is not related with the Cygnus Loop.

There are mainly two types of mechanisms that can cause depolarization: depth depolarization and beam depolarization~\citep{sokoloff+98}. For depth polarization, the synchrotron-emitting and Faraday-rotating medium coexist, so that polarization from different locations along the line of sight experiences different Faraday rotation. Thus adding these emission components partly or completely cancels the polarization. For beam depolarization, the Faraday-rotating medium is in front of the synchrotron-emitting medium. Because of the RM fluctuation, the polarized emission varies across the beam. Averaging the emission within the beam also reduces the polarization following $\Pi=\Pi_0\exp(-2\sigma_{\rm RM}^2\lambda^4)$, where $\Pi$ and $\Pi_0$ are observed and intrinsic degree of polarization, $\sigma_{\rm RM}$ is the dispersion of RMs, and $\lambda$ is the wavelength. Beam depolarization is the primary cause of the mottled structures in polarization images~\citep[e.g.][]{sun+14}, and has been found to account for depolarization towards repeating fast radio bursts~\citep{feng+22}. 

\begin{figure}
    \centering
    \includegraphics[width=0.8\textwidth]{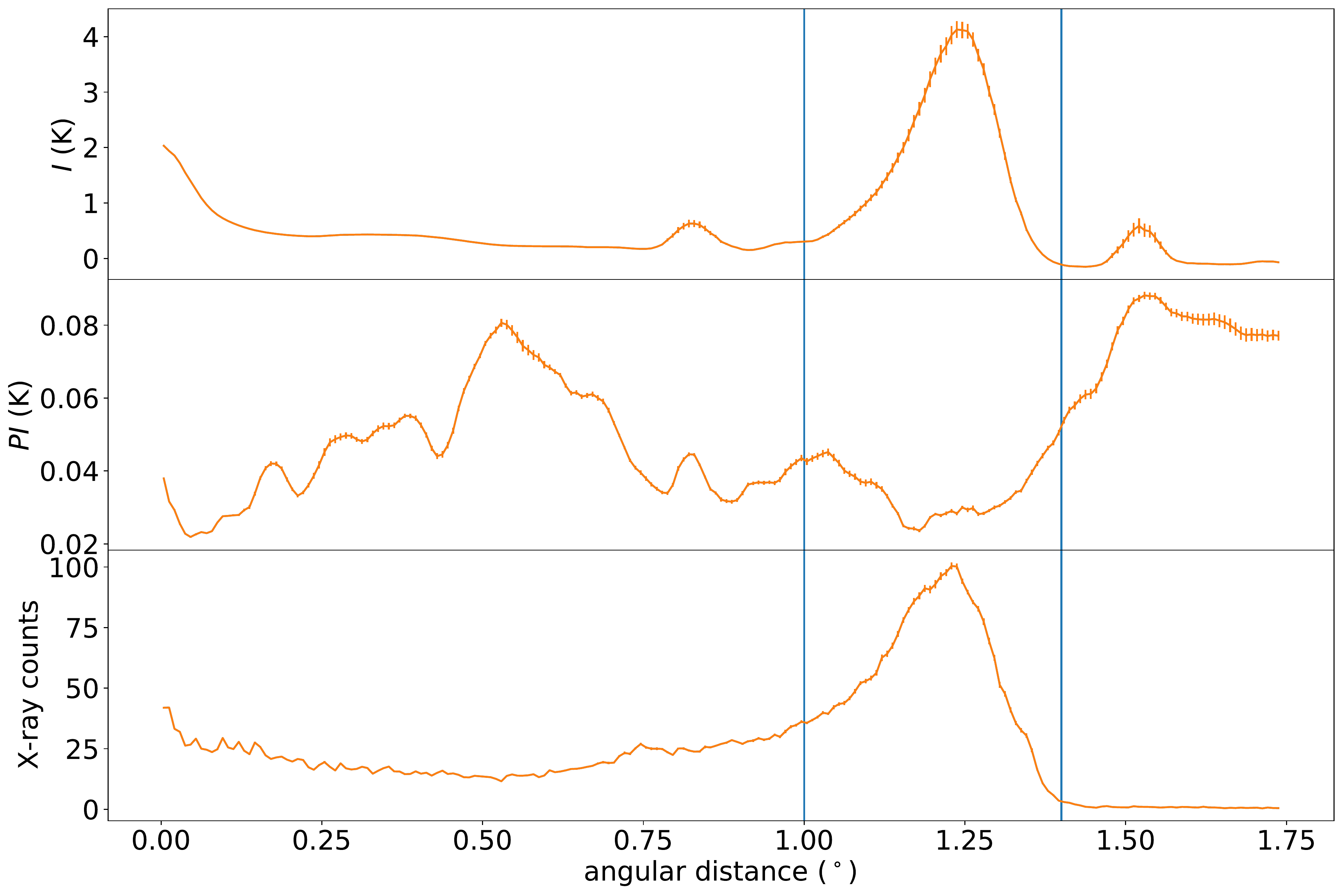}
    \caption{Radio total intensity ($I$), radio polarized intensity ($PI$), and X-ray intensity versus angular distance from the reference position $(20^{\rm h}51^{\rm m}36^{\rm s},\,31\degr03\arcmin)$ for the northeastern shell. The vertical lines indicate a radius of $1\degr$ and $1\fdg4$, respectively.}
    \label{fig:prof}
\end{figure}

In Fig.~\ref{fig:prof}, we show the profiles of radio total and polarized, and X-ray intensities versus angular distance from the reference position $(20^{\rm h}51^{\rm m}36^{\rm s},\,31\degr03\arcmin)$ which is approximately the geometric center of the northern part of the Cygnus Loop (Fig.~\ref{fig:cyg_i}). The profiles were derived towards the northeastern shell. The radii of $1\degr$ and $1\fdg4$ are also marked in Fig.~\ref{fig:prof}. Within this annulus of about $0\fdg4$, both radio total and X-ray intensities are high, but the polarized intensity is low. Outside the radius of $1\fdg4$, there are virtually no radio total and X-ray intensities, but there is considerable amount of polarization that originates from the interstellar medium of the Galaxy rather than from the Cygnus Loop. 

The polarization patches adjacent to the Cygnus Loop~(Fig.~\ref{fig:cyg_pi}), which cause the increase of polarized intensity beyond the radius of $1\fdg4$ (Fig.~\ref{fig:prof}), imply that the Galactic polarized emission behind the Cygnus Loop is smeared out. This favors the beam depolarization as the cause of disappearance of polarization towards the northeastern and western shell. The beam depolarization results in mottled polarization structures which are also prevalent across the SNRs. The ratio of the degree of polarization between 2.7~GHz and 4.8~GHz is about 80\%~\citep{sun+06}, meaning the RM fluctuation is about $\sigma_{\rm RM}\sim30$~rad~m$^{-2}$. This amount of RM fluctuation is sufficient to cause complete depolarization at 1.4~GHz. 

From the composite H$\alpha$ all-sky map~\citep{finkbeiner+03}, the H$\alpha$ intensity ($I_{{\rm H}\alpha}$) is about 30~Rayleigh towards the northeastern shell, where only low angular-resolution ($\sim 1\degr$) data from Wisconsin H$\alpha$ Mapper~\citep{haffner+03} are available. Following \citet{haffner+98} and neglecting absorption, the emission measure (EM) can be derived as ${\rm EM}=2.75T_4^{0.9}I_{{\rm H}\alpha}$. Here EM is in pc~cm$^{-6}$, the electron temperature $T_4$ is in 10$^4$~K, and $I_{{\rm H}\alpha}$ is in Rayleigh. Taking $I_{{\rm H}\alpha}$ of 30~Rayleigh and the typical value of $T_4=0.8$, we obtained an EM of 67.5~pc~cm$^{-6}$. The brightness temperature contributed by the thermal gas of temperature $T=8000$~K at frequency $\nu=1.28$~GHz can be estimated as $8.235\times 10^{-2} T^{-0.35}\nu^{-2.1} {\rm EM}\approx0.1$~K~\citep[e.g.][]{sun+11}, which only takes up several percent of the total intensity. The radio emission at 1.28~GHz is predominately from synchrotron emission.

We used the distance of $725\pm15$~pc for the Cygnus Loop, which was derived from the parallaxes of the stars associated with the SNR~\citep{fesen+21}. The width of the northeastern shell is about $0\fdg4$~(Fig.~\ref{fig:prof}), corresponding to a size of 5~pc. Assuming a spherical shell, the path length along line of sight, $L$, is thus about 25~pc. 

Towards the northeastern shell, the thermal electron density $n_e$ can be derived according to ${\rm EM}\sim n_e^2L$, which is about 1.6~cm$^{-3}$. Here the volume filling factor for the thermal electrons is assumed to be 1. The RM fluctuation can be represented as $\sigma_{\rm RM}\sim 0.81 n_e b \sqrt{Ll}$~\citep[e.g.][]{sun+11}, where $b$ represents the strength of random magnetic field in $\mu$G, and $l$ is the correlation scale for magnetic field fluctuations in pc. For beam depolarization, $l$ should be much smaller than the beam width of $4\arcmin$ or 0.8~pc. Therefore the lower limit of $b$ is about 5~$\mu$G. Towards the western shell, where the H$\alpha$ data are from the high angular resolution ($\sim 1\arcmin$) Virginia Tech Spectral line Survey~\citep{dennison+98} , the H$\alpha$ intensity is in the range of 50 -- 140~Rayleigh, corresponding to EM of 112.5 -- 315.0 pc~cm$^{-6}$, and electron density of 2.1 -- 3.5~cm$^{-3}$. This leads to an estimate of the lower limit of $b$ of about 3~$\mu$G. 

The average RM derived from the data at 4800~MHz and 2695~MHz is about $-70$~rad~m$^{-2}$ towards the northeastern and western shell~\citep{sun+06}. The intrinsic RM is thus about $-$50~rad~m$^{-2}$ after taking into account the foreground contribution of about $-20$~rad~m$^{-2}$. The strength of the regular magnetic field can then be estimated to be about 1.5~$\mu$G, which is smaller than that of the random magnetic fields. 

\citet{tutone+21} modeled the emission of the Cygnus Loop at $\gamma$-ray, X-ray, UV and radio bands, and obtained a magnetic field of about 10~$\mu$G towards the northeastern part. For the western part, they used a magnetic field about 244~$\mu$G derived by \citet{raymond+20} towards thin optical filaments. \citet{loru+21} modeled the $\gamma$-ray and radio emission and obtained a magnetic field of about 10~$\mu$G for the whole SNR. Note that these magnetic fields include both regular and random components. Together with these previous results, our estimate with a regular field of 1.5~$\mu$G and a lower limit of 5~$\mu$G towards the northeastern part and of 3~$\mu$G towards the western part for the random field corroborates that the random field is much larger than the regular field.

Towards the southern part, the H$\alpha$ intensity is at a similar level to the northeastern shell, but there is nearly no depolarization. This suggests that the magnetic field fluctuation is negligible. 

\subsection{A multi-scale analysis}
\begin{figure}
    \centering
    \includegraphics[width=0.95\textwidth]{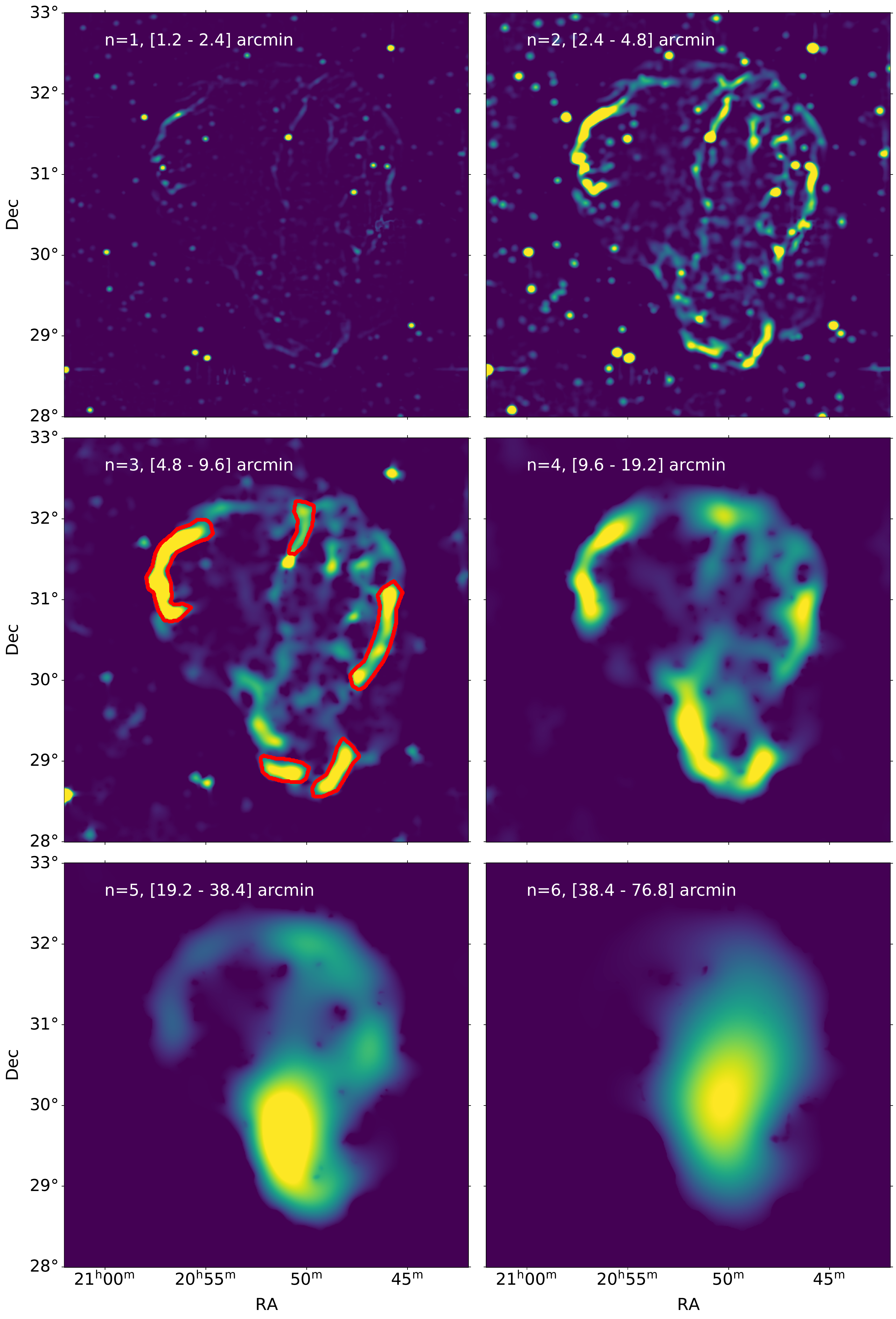}
    \caption{Component images from the constrained diffusion decomposition (CDD) of the Cygnus Loop. The component number $n$ and the corresponding range of the angular scales are marked in each panel. The areas used for TT-plots are outlined by red polygons. The color scale extends from 0 (dark) to 0.8~K~$T_{\rm b}$ (bright).}
    \label{fig:cdd}
\end{figure}

We decomposed the total-intensity image shown in Fig.~\ref{fig:cyg_i} into components of various angular scales using the constrained diffusion decomposition (CDD) method developed by \citet{li+22}. In contrast with other multi-scale decomposition methods, the CDD method does not produce artifacts containing negative values around extended structures. The component images are shown in Fig.~\ref{fig:cdd}. For component $n$, the corresponding scales of the structures are larger than $2^n\Delta$ but smaller than $2^{n+1}\Delta$, where $\Delta=0\farcm6$ is the grid size of the map. 

Both the northern and southern parts of the Cygnus Loop show bright shell structures predominantly in components $n=2$ and $n=3$~(Fig.~\ref{fig:cdd}). For component $n=2$, the angular scale is roughly equal to the beam size, and the structures in this component might be susceptible to beam averaging. We will focus on component $n=3$ in the analyses below. The northern part consists of the northeastern shell (NE), the central filament (C), and the western shell (W); whereas the southern part can be decomposed into two shells at southeast (SE) and southwest (SW), respectively. We outlined these features in Fig.~\ref{fig:cdd}. 

\begin{figure}
    \centering
    \includegraphics[width=0.95\textwidth]{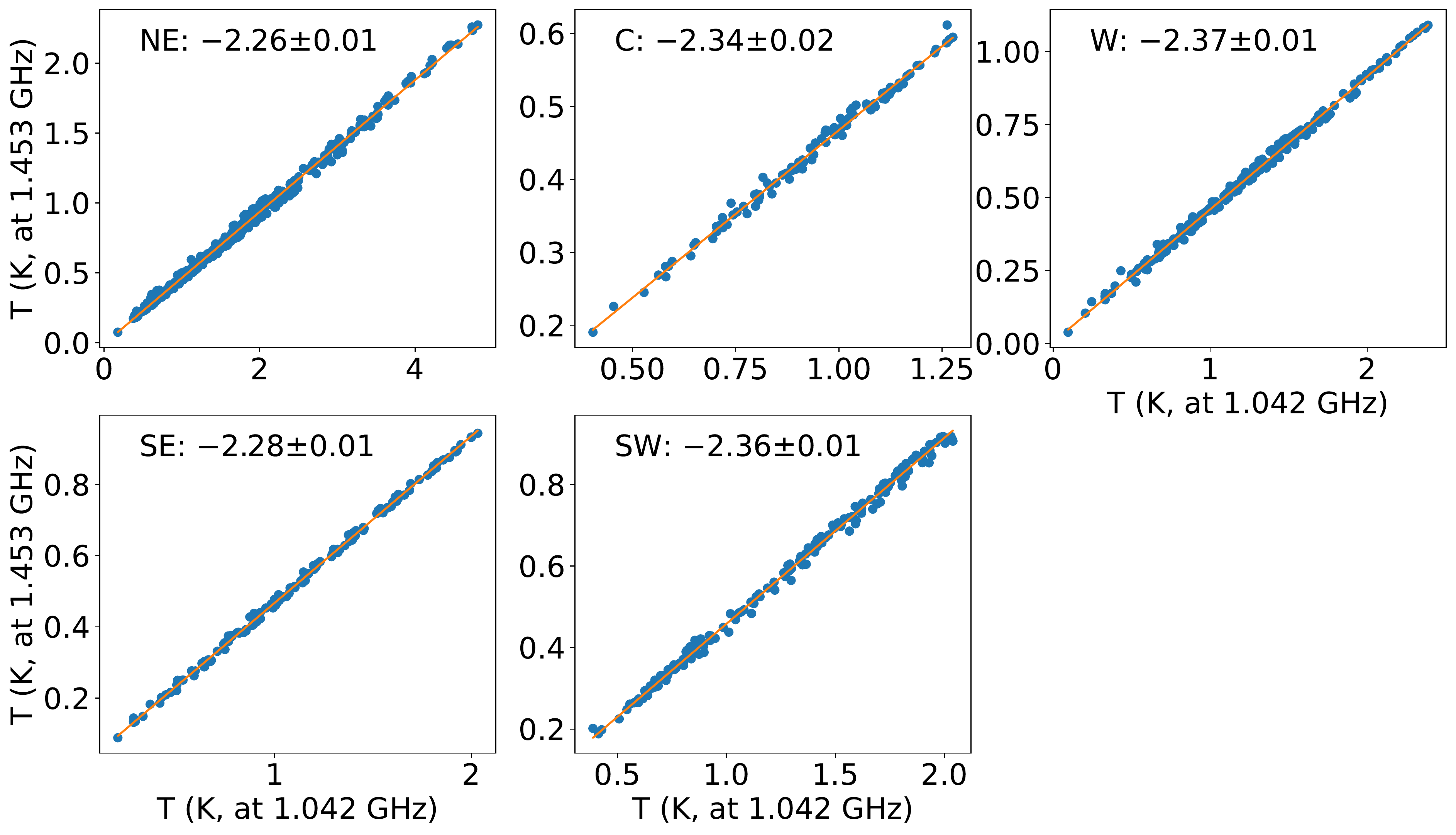}
    \caption{TT-plots for the shell structures in component $n=3$.}
    \label{fig:tt_cdd3}
\end{figure}

We made temperature versus temperature plots \citep[TT-plots,][]{turtle+62} to determine the brightness temperature spectral index $\beta$, defined as $T_\nu\propto \nu^\beta$. The flux density spectral index $\alpha$, defined as $S_\nu\propto\nu^\alpha$ with $S_\nu$ being the flux density, can then be obtained as $\alpha=\beta+2$. We picked up the two frequency channels well apart at 1.042~GHz and 1.453~GHz to derive TT-plots for component $n=3$, and the results are shown in Fig.~\ref{fig:tt_cdd3}. Note that the features outlined in Fig.~\ref{fig:cdd} for component $n=3$ have a very good correspondence between these two frequencies.

For component $n=3$ with angular scales from $4\farcm8$ to $9\farcm6$, the northern and southern parts of the SNR have very similar spectra. The spectral indices of the northeastern and southeastern shells are both around $-2.27$, and the spectral indices of the western and southwestern shells are both around $-2.36$. 

\subsection{Remarks on the morphology of the Cygnus Loop}

The northern part of the Cygnus Loop has been interacting with dense interstellar clumps, which can be inferred from the multi-wavelength view of the SNR. The interactions can increase the random magnetic field which causes complete depolarization towards the northeastern and western shells at 1.28~GHz. The similarities of foreground RMs and brightness-temperature spectral indices further suggest that these two parts belong to the same SNR.

\section{Conclusions}
\label{sect:conclusion}

We obtained total-intensity and polarized-intensity maps of the Cygnus Loop by combining observations scanning in RA and Dec directions. We also derived an RM map using RM synthesis. 

We compared the properties of the northern part with the southern part of the Cygnus Loop in several aspects. From a multi-wavelength view, the interaction with the ambient interstellar clumps caused the northern part to resemble an SNR in the radiative phase, whereas the southern part resembles a typical SNR in the adiabatic phase. The depolarization in the northern part indicates that the random magnetic fields are stronger than the regular fields. In contrast, the random fields are negligible in the southern part. We decomposed the total intensity into components of various angular scales. The TT-plots for the structures in different component maps show that the northern part and the southern part have similar spectra. These suggest that the northern and southern parts belong to the same SNR.

\begin{acknowledgements}
XS is supported by the Cultivation Project for FAST Scientific Payoff and Research Achievement of CAMS-CAS. XS and XL are supported by the Science \& Technology Department of Yunnan Province - Yunnan University Joint Funding (2019FY003005). XS and XG are supported by the CAS-NWO cooperation programme (Grant No. GJHZ1865). We would like to thank Dr. Guangxing Li for fruitful discussions and Dr. Patricia Reich for careful reading of the manuscript. We also thank the anonymous referee for the comments that help improve the paper.
\end{acknowledgements}

\bibliographystyle{raa}
\bibliography{bibtex}
\end{document}